# Hydromechanical Modelling of Carbon Sequestration in Sedimentary Rocks

Hamed.O.Ghaffari & Mamadou Fall

*Dept. Civil Engineering, University of Ottawa, Ottawa, Ontario, Canada*

**Abstract**:

In this study, over different scenarios we will simulate a week coupling of hydromechanical loads in a long term $CO_2$ injection with a hypothetical reservoir while the effect of pore water pressure and then multi-phase flow procedure has been ignored. In the first basic case the homogenous case has been considered when the theory of poroelasticity was employed. Second case covers the effects of directional heterogeneity, constructed by random faults, on the flow paths of gas and other attributes of the system. Also, in the latter case the impact of stress state as an active loads (body loads) has been regarded. Thanks to multiple directional heterogeneity, which induces only one heterogenic parameter (intrinsic permeability), distinguishable flow paths can be recognized. In another process, the failure ability of system regard to Mohr-Columb criterion is measured as well as options that, presumably, the system has continuum faults (zero cohesion). The results over different cases shows absedince of ground surface (heave), more probable propagation of failure area and the role of directional heterogeneity to change the evolution path of system.

# 1. Introduction

Carbon dioxide sequestration, storage and capture (CCS), is one of the main options to reduce the gas emissions which proposed nearly 10 years ago. Due to effect of carbon emissions on the general climate change CCS is one of the main situations to decrease the climate change rate. Selection of reservoirs, analysis and considering essential short and long term effects of injection gas within the formation needs to more focus in multi-disciplinary fields [1],[2].

Differences of gas injection in short term against long term gas capturing induces the high rate of injection (millions ton per year) which lays on dynamic transferring of pressure and shock wave of gas plus high risk of crushing of formation. This later consideration will be more catastrophic where some or collected of unknown fractures are within the formations or are inducing by the procedure so that may connect and intersect with some narrow pathways. The effect of High temperature (50-80 $^c$ below 800-1200 underground), pressure of formation and chemical component of water dramatically change the behaviour of carbon sequestration [3] while all of these forces are coupling with each other and driving the system.

In another view, considering of gas plume within the other fluid induces two (or multi) phase flow which due to intrinsic heterogeneity of the formations will follow different regimes of flow. Generally, complexity is more than this especially when we consider the trapping mechanism of injected gas that is an index of evolution of gas saturation related to time and space which change the models parameters especially ones are related to the two-phase flow. Analysis of all of the mentioned complexities within the system is more complicate in mathematical language and needs, as well as other physical events, to simplification. Different authors and researches have focused on hydraulic properties of gas plume motion and recently the effects of chemical components of in situ water on the properties associated with the effects of thermo- hydromechanical forces is under attack [4],[5]. In this way, Recognizing of weak and strong coupling over parameters plays an outstanding role in analysis of the system such permeability depend on stress and temperature or capillary pressure strength dependency to stress and chemical component of pore water[6] .

In this study, we will focus on the hydromechanical behaviour of the injected gas when with poroelasiticty theory and Darcy law the gas evolution within a drayed area will be analysed . In this way, two basic cases with and without passive heterogeneity using finite elemnt method will be analysed. The effects of directional heterogeneity, constructed by random faults, on the flow paths of gas and other attributes of the system. Thanks to multiple directional heterogeneity, which induces only one heterogenic parameter (intrinsic permeability), distinguishable flow paths can be recognized. In another process, the failure ability of system regard to Mohr-Columb criterion will be measured as well as options that, presumably, the system has continuum faults (zero cohesion). The results over different cases shows absedince of ground surface (heave), more probable propagation of failure area and the role of directional heterogeneity to change the evolution path of the system.

## 2. Governing Equations

In this part, we will present the general formulation of week coupling of two phase flow and mechanical loads which is associated with poroelasticity theory. In week coupling some aspects of the solution is ignored. This can be inferred from realistic week coupling over the regarded agents (forces-loads) or complexity reduction of system, i.e., physical simplification assumption will be added to the other initially assumptions. This study will not cover coupling of porosity (then permeability) and stress –stress dependent permeability. Also, in a particular simplification, we will not consider the effect of pore water pressure .Then the presented equations will be solved (by Finite Element Method) with the effect of one phase in fluid part which is going to couple with mechanical deformation upon the gradient pressure. Simultaneously, the gradient of fluid pressure plays the role of body load force in mechanical deformation term (force equilibrium equation), scaled by Biot's coefficient (here, compressive stress has minus sign).

The fundamental equation for mechanical physics is based on the force equilibrium equation [7]:

$$-\nabla.\sigma = F \quad (1)$$

where $\sigma$ is the total stress (tensor) and $F$ is the external or body force. To insert the effect of two phase flow pressure, the general solution comes from the mixture pressure, states the mean pressure of fluid as the weighted mean of pressure based on saturation of each phase. It can be expressed by Eq.2:

$$\overline{P} = S_w P_w + S_{nw} P_{nw} \quad (2)$$

in which $S_w$, $S_{nw}$, $P_w$ and $P_{nw}$ are saturation and pressure of wetting and non wetting phase, respectively. For single phase flow, $S_{nw}$ and $P_{nw}$ ignored and $\overline{P} = P_w$. Similarly the density of mixture fluid can be followed by:

$$\overline{\rho} = S_w \rho_w + S_{nw} \rho_{nw} \quad (3)$$

Then total stress due to effective stress ($\sigma'$) and fluid pressure is expressed as:

$$\sigma = \sigma' + \alpha \overline{P} \quad (4)$$

where the parameter $\alpha$ is called Biot-Willis coefficient. Implementation of pore pressure in Eq.1, we will have:

$$-\nabla.\sigma = F - \alpha\nabla\overline{P} \qquad (5)$$

For single phase flow (which is our case), the simplest case: Darcy empirical law is considered:

$$v = -\frac{k}{\mu}(\nabla p - \rho g\nabla z) \qquad (6)$$

where $\mu$ is the dynamic viscosity of the fluid and $k$ is the permeability of rock formation. As we mentioned fluid pressure can affect the stress state in the porous media. On the other hand, deformation of porous media can affect the fluid flow through formations. Then the coupling Partial Differential equations based on poroelasticity theory are as follows [7]:

$$G\nabla^2 u + (K_d + \frac{G}{3})\nabla.(\nabla u) = \alpha\nabla\overline{P} \qquad (7)$$

$$\frac{1}{M}\frac{\partial \overline{P}}{\partial t} - \nabla.(\frac{k}{\mu}(\nabla\overline{P} - \overline{\rho}_f g\nabla z)) = -\alpha\frac{\partial(\nabla u)}{\partial t} \qquad (8)$$

where $G$ is shear modulus of solid part, $K_d$ is the drained bulk modulus and $u$ is displacement. $M$ is the Biot's Modulus is defined as [8]:

$$\frac{1}{M} = \frac{\alpha - \phi}{K_s} + \frac{\phi}{K_f} \qquad (9)$$

in which $K_s, K_f$ are bulk modulus of solid and the fluid. Generally, the Biot's Modoulus can be expressed by storage coefficient ($\frac{1}{M} = S_\alpha$) for each formation. Another coupling (It has been ignored in this study), is dependency of porosity (permeability) to stress. For this case, some empirical and semi-analytical expressions have been mentioned in the literature especially change of permeability as a result of volumetric strains and Rutqvist and Davis functions [9], [10].

To evaluate the possible failure area within the formation in post processing step, a well known Coulomb failure criterion will be used [11]:

$$fail = (\sigma_3 + P) - (\frac{1+\sin\phi}{1-\sin\phi})(\sigma_1 + P) + \frac{2\cos\phi}{1-\sin\phi}C(1+\frac{\sigma_2 - \sigma_1}{\sigma_3 - \sigma_1}) \qquad (10)$$

in which $C$ is the Coulomb cohesion and $\phi$ is the Coulomb friction angle. *fail = 0* indicates the onset of rock failure; fail < 0 denotes catastrophic failure; and fail > 0 predicts stability [11]. Consider that this criterion can be used to insert active growth of dislocation or heterogeneity such join in each point (area/ node). Then such growth/decaying (healing) procedure based on the similar criteria will change other properties of rocks such porosity and this will affect to permeability and capillary pressure in two phase flow cases. We will discuss on this topic in the next section.

## 3. Computational Model: Homogeneous and Heterogeneous cases

In this part, the modeling results over two basic cases will be presented. It must be reminded all the presented models ignore the fully hydraulic cases, i.e. the role of water (or brine effects) on the flow behaviour of $CO_2$ and the effects of pressure, temperature and salinity dependence of $CO_2$ dissolution in the aqueous phase have been neglected. Figure 1 shows the considered system with three different rock formations. The properties of the formation related to the governing equations can be followed at Table 1. Formation3 is the reservoir (or aquifer) with high permeability and weak strength parameters while the formation2 can be presumed as the cap rock with low permeability which is the necessary part of an area to be appropriate for sequestration of gas.

The thin layer in the ground surface of the system completes our model. For hydraulic boundaries all of boundaries except ground surface (and thin layer borders) have no flux. The ground surface and boundaries of first layer, the pressure is assumed zero. For The boundary condition for mechanical part, we allow normal deformation for left hand because it is near to injection point when right hand is mixing with fixed and rollers. The roller part is more related to deformation of ground surface. The gas will injected with the rate of 600 Pa. per year. In practical case, this value is so higher than this value, say, nearly 7000pascal per hour. However, without losing generality, simulation over more years can give a scaled value related to realistic case. It means in our model we expand the time when the space is same. Two basic cases were considered. In the first situation, the formations are homogenous and the parameters of the formation are as well as table 1. Also, to evaluate the susceptibility of media to slip due to implicitly considering of fault, the value of cohesion strength gets zero. In other word, we assume each node of media can be assumed to has an unknown fault property, i.e., implicit (continuum) regarding of fault.

In the second case, the system accepts some directional random heterogeneity, which is created by passive change (offline) of rock's formation. In this case, for each generation of fault, the permeability of intersected rocks is decreased by a coefficient when other properties of the formation (also, C=0) are preserved. In another view, the formations within the system over the successive times (in an inactive way: without coupling with stress or other criterion) catch a random fault system. For the mentioned cases, we didn't consider the gravity (or initial stress sate) body loads. However, for the last case, the body force plus initial stresses were considered. In other word, starting of gas injection the stress of the system due to gravity and gas pressure will change. This means simulation of an unconsolidated heterogeneous system simultaneously is undergoing the gas injection effect. Maybe a realistic case in nature can be mimicked by considering a hypothetical volcanic activity which is acting on a new finished activity (then yet, has not been consolidated).

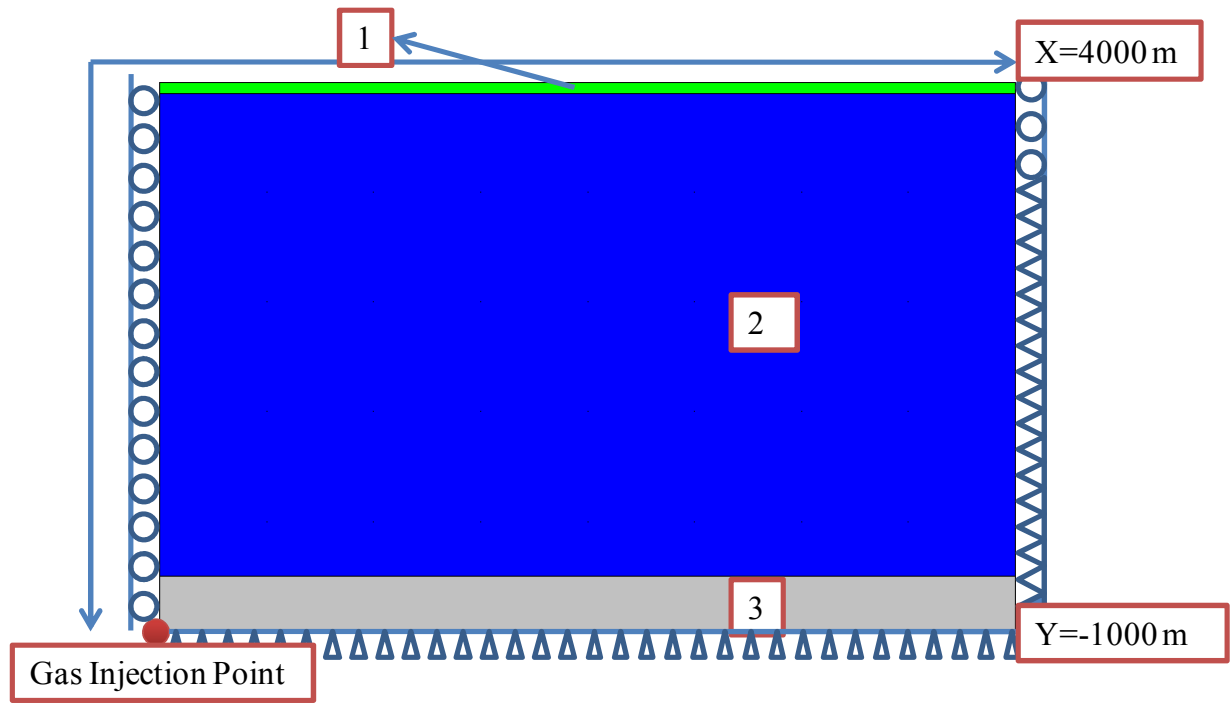

Figure 1.Vertical profile of the model with boundary conditions and initial rock formations

Let us start the interpretation of the results with first case. Figure 3, shows the variation of displacement, pressure and failure value for first case. As It ahs been depicted the pressure of gas in the reservoir is much higher than the other parts and so, with time passing It will be transferred along the high permeable area. The injection point area has the highest gas pressure as we expected. As one can follow due to disconnection of the reservoir area, the injected gas will increase the stress (compressive stress) and then displacements (heave-absedince) of the system. Figure 3b shows that in the locked reservoir with low level of strength properties; the injected pressure after 1000 years ( $c \neq 0; without\ initial\ stress\ state$ ) induces high risk to initiate failure. By following boundary condition at left hand, and comparing with left hand displacements (Figure3d), It cab understood that highest deformation is in the gas injected point. In figure 4 we plotted the evolution of three point located at reservoir, cap rock and ground surface.

As we mentioned the failure in reservoir has highest possibility when after nearly 4000 years the failure will be started 9Figure 4d). The asymptotic behaviour of gas velocity within reservoir has been plotted in figure 4d which shows at the middle of reservoir the velocity of gas (after quasi-phase transition in log-log coordinate) goes to 4mm/s after 10000years. Also, Figure 4c proves the using of mixing roller and fixed boundary condition while foe first case the displacement depression and freedom will be higher than the second.

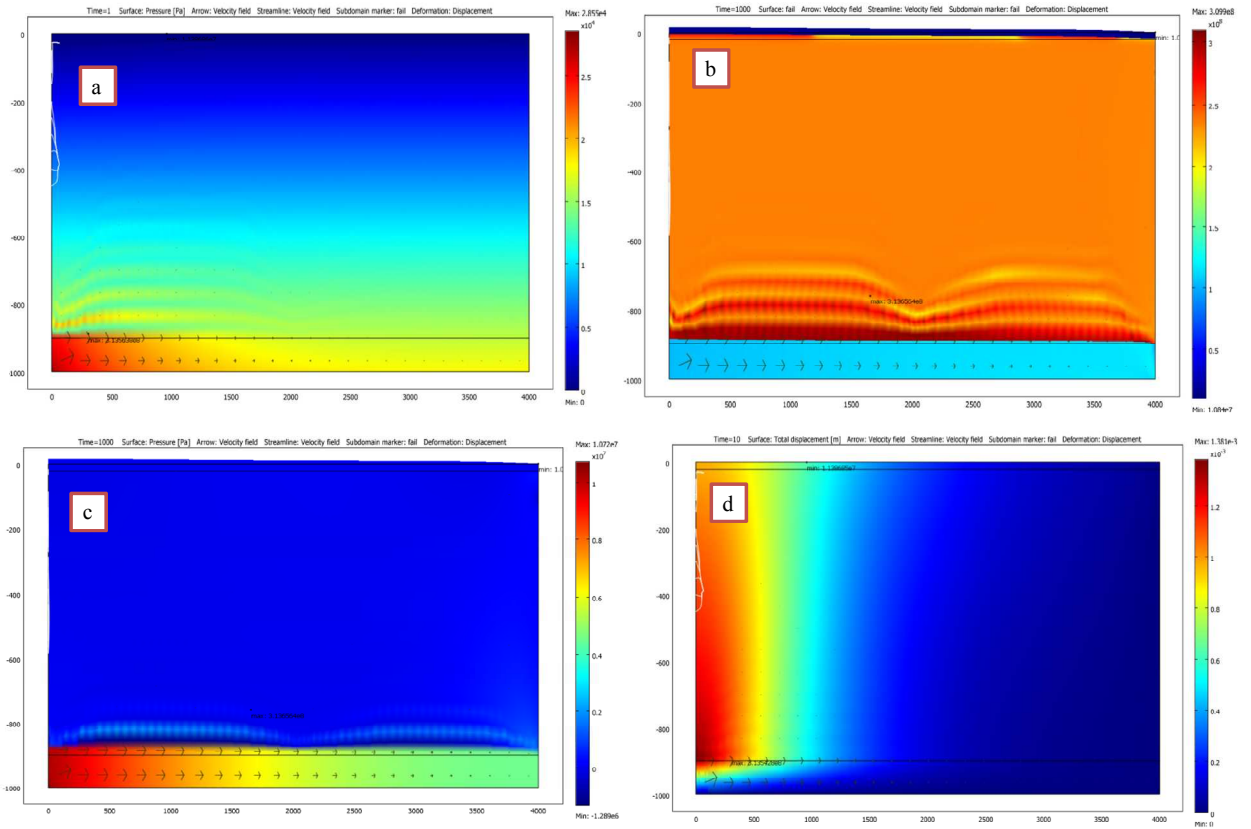

**Figure 2.** variation of displacement, pressure and failure value for first case: a) pressure of gas after one year; b) failure value after 1000 years; c) pressure after 1000 years and d) absolute displacement after 10 years of gas injection.

Table 1. Properties of the formation in the model

| Material properties | reservoir | Cap rock | Ground surface |
|---|---|---|---|
| E(Gpa) | 5 | 5 | 5 |
| $\nu$ | .25 | .25 | .25 |
| $\rho$ (density kg/M$^3$) | 2260 | 2800 | 2000 |
| $\phi$ (friction angle) | 25 | 25 | 25 |
| K(permeability) | $3e-11$ | $3e-16$ | 3e-10 |
| S(storage coefficient) | $1e-6$ | $1e-5$ | 1e-6 |
| C(cohesion strength-MPa) | 20 | 50 | 3 |

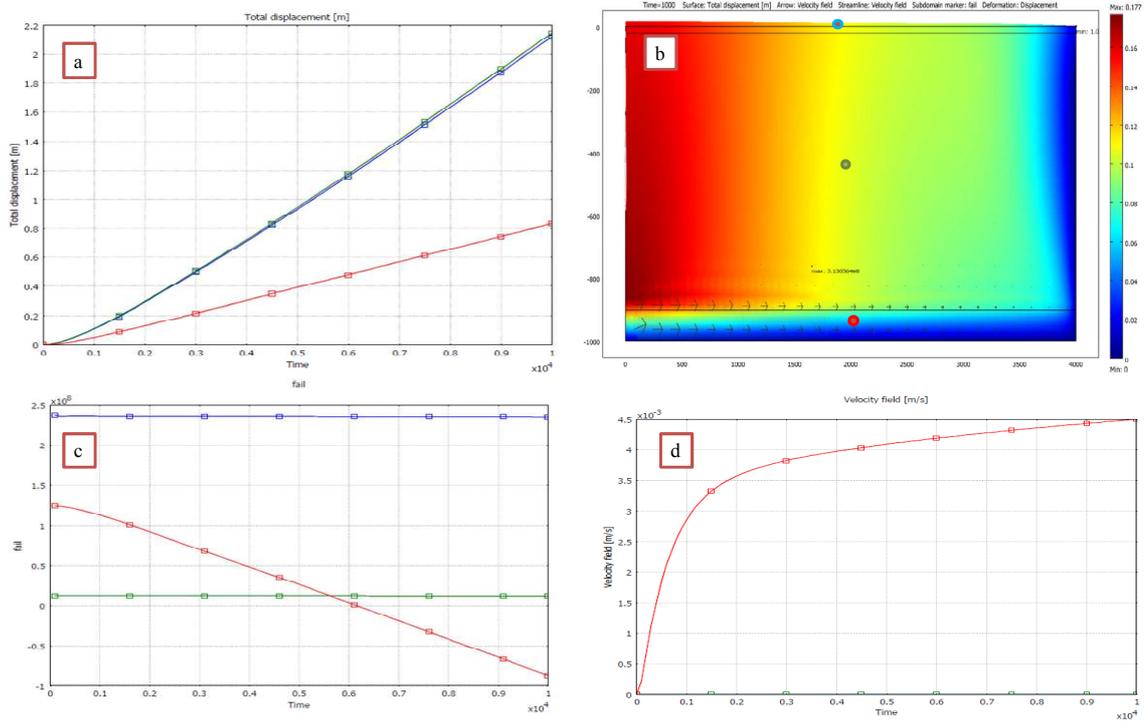

**Figure 3.** First case a) Total displacement along three different points in reservoir, cap rock and ground surface; b) Total displacement after 1000 years ;c) change of fail value for three points along time and d) velocity of gas within the middle of reservoir along 10,000 years.

Now consider the case that all of the area has possibility to has fault then has ability to slip and we set c=0. The directed conclusion can be inferred from figure 4, in which total displacement (absolute value) and failure values along a cross-section have been illustrated. As we can see the pressure of gas for this case affects the failure value while the failure after 3000 years within aquifer is much higher than the previous case (fail in the same time is nearly 4e08 against -1e07). Consider in this case after 500 years the ground surface will have nearly 10 cm heave and uplifting. Again, the boundary condition can be followed in the displacements of right and left walls.

The next case is related to the collection of large randomly directed faults. For this aim, we completed a simple program to generate passive faults while the length and dip of generations are changing based on a uniform distribution probability function. As if we know that the length of joints generally is obeying from a power law however our aim was to analysis the effect of one-dimensional heterogeneity to gas flow and failure risk. For each generation (assume each fault generation is related to one geological period), the permeability (and only permeability) will decreased with a coefficient. Intersected points between faults are more close to a collapsed area and then depend on number of intersections the permeability will decreased much more than the former case.

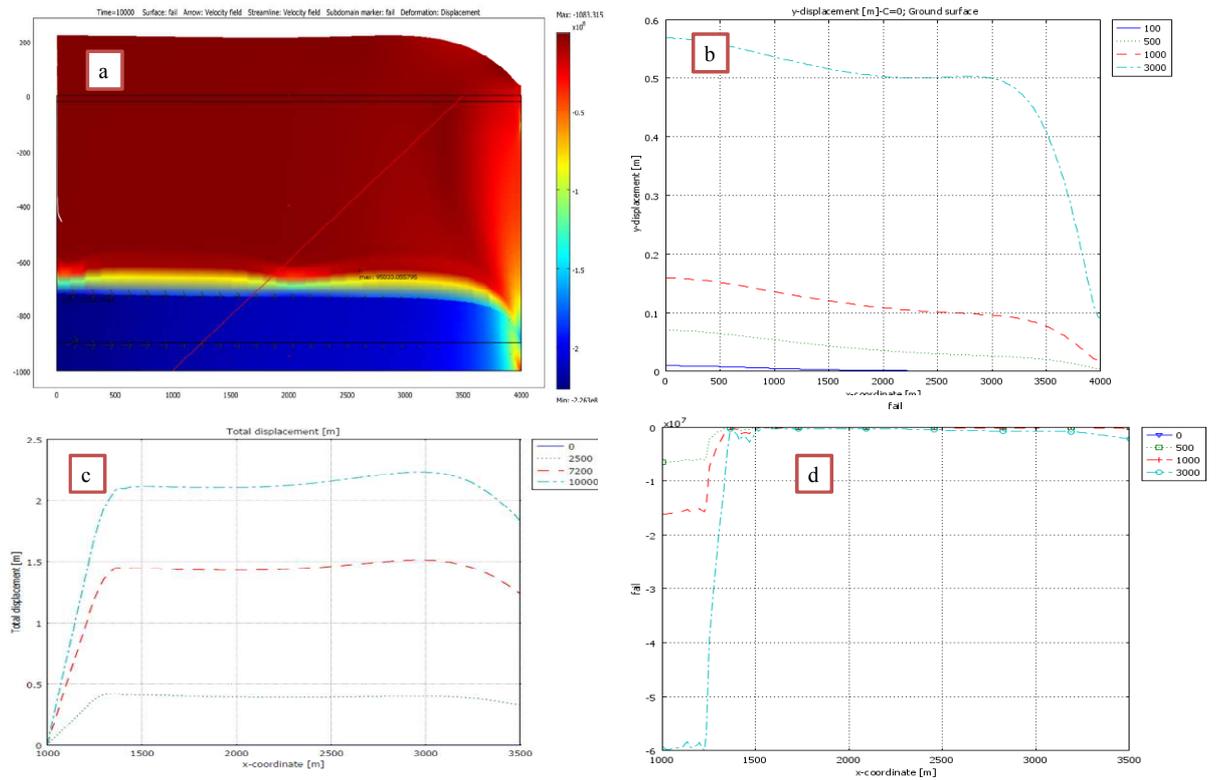

**Figure 4. First case, C=0: a) fail value after 10000 years; b) Displacement of growth surface along different times ;c) Displacement over the highlighted line at (a) and d)fail value over the highlighted line which crosses three formations .**

Different models have been presented to describe the evolution of a rock joint (and in big scale faults) [], []. In other word, using the basic concepts such stress state (Normal or/and shear stresses), rock joint properties (aperture, roughness, stiffness, so on), the active behaviour of fault/rock joint is modeled. However, growth, propagation of joints and then damage for a long term must be considered to complete the small scale view and local evolution of rock joints. We have neglected this event in our simulation. In Figure 5 a and b, we have shown the fault generation over 500 and 3000 times respectively. In the following of this study, we will use second generation.

Comparison of failure value after 1000 years from the gas injection with the changed permeability values (figure 5 c and 7c) shows how without changing the strength properties of rock material and with interaction of gradient pressure and deformation (and change of stress state –induced stress) ; the failure distribution will change. As the former cases the gas injection point is the place with high probability to fail. However the failure distribution coincides with the permeability values where the congested area of faults has big chance to fail. In figure 6a, the distribution of gas after 1000 year s has been illustrated while follows a similar pattern of directional dense faults.

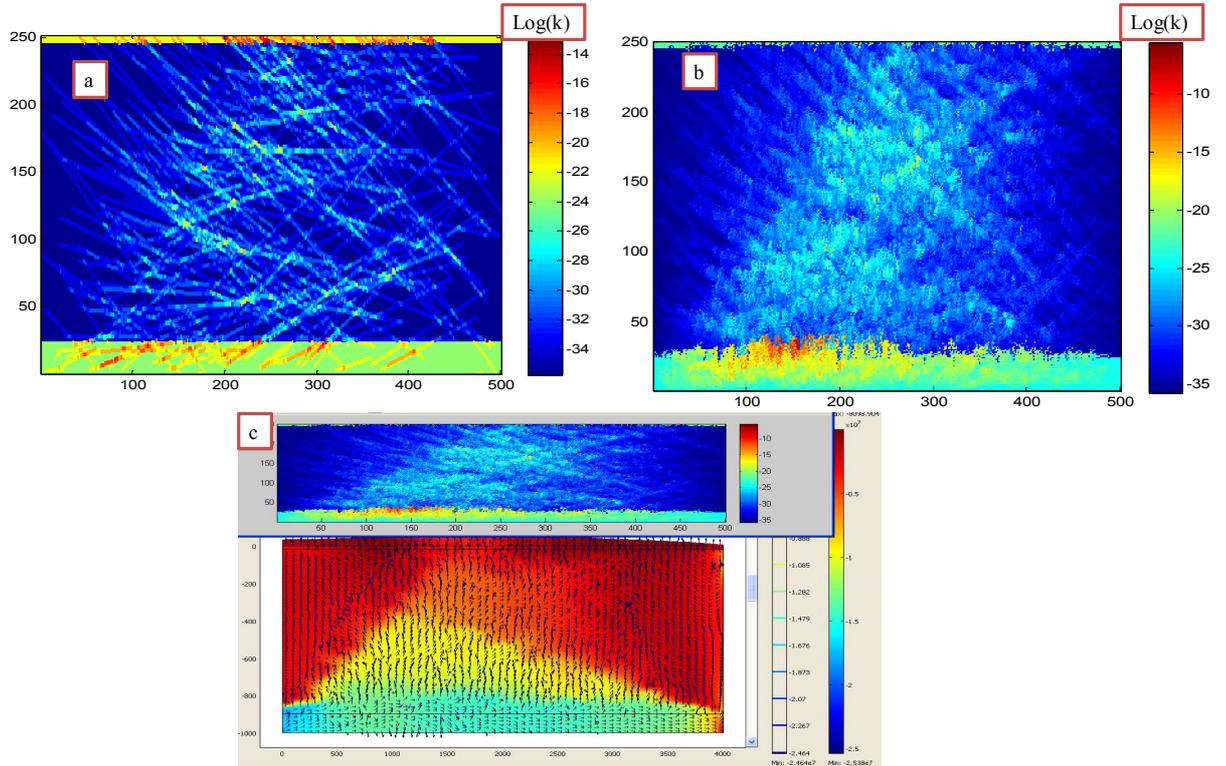

**Figure 5.** Generation of random directional heterogeneity to change of permeability: a) 500 generations; b) 3000 times generations and c) Comparison of generated faults area with the fail value (after 1000 years) of the model.

Consider high permeable zones (hub positions of faults) controls the attributes of system which means those points (for example see the crushed zone within the reservoir) conduct the flow of gas. In figure 6b and c we have depicted the velocity of gas after 10 and 500 years which are following with deformation counters. The gas paths also are followed by pattern of permeability distribution. Consider that the percolation points of gas –from reservoir towards ground surface- are on the high dense places of intersected faults. In other word, the hubs of connections (random intersection) determine the future pathway of gas flow. In our case these crushed areas are more concentrated in the interface of reservoir and cap rock(x=500-1200)-see so figure 7d.

With presenting the failure values of ground surface, it was determined despite the first case; the failure is not nearly uniform and is governed by direction of heterogeneities. For example, after 500 years the failure value at x=1500 is 4 times lower than the other points (Figure 7a). This is particularly is matched with the variation of ground surface displacements (Figure 7b and c). Referring to failure profiles in ground surface and considering time evolution of deformation, it appears that in a long terms after gas injection (due to orientation and induced heterogeneity) the non-uniformity of deformation will be increased (figure 7a). It must be reminded in the aforementioned cases we have not considered any initial or active stress field.

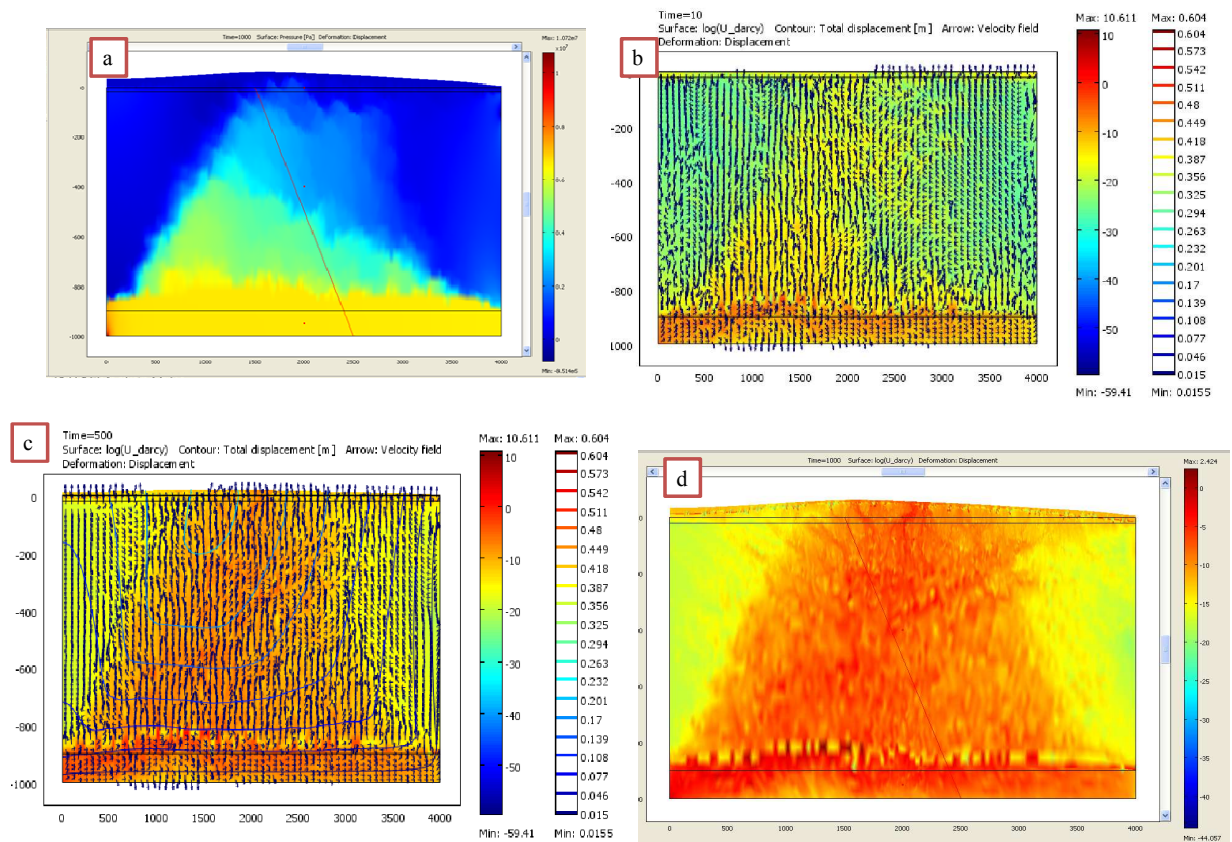

**Figure 6.a) Pressure of gas after 1000 years ;b) logarithm of gas velocity and normalized direction of gas flow paths after 10 years) The evolution of flow paths and gas velocity plus deformation counters after 500 years and d) gas velocity (scaled in logarithmic way) after 1000 years**

Now we consider initial stress field (compressive stress field based on gravity loads : $\sigma_x = \sigma_y = -\gamma z$) which is accompanied with active stress effects as body forces in right hand of Eq.5. Interpretation of such model can be imagined as simultaneously gas injection and consolidation of the formation which may is more related to volcanic activity. For a long time, the gravity loads frustrate the effects of gas pressure .However, for a uniform active gas injection after nearly (300-400 years) the gas velocity field (and paths) are following the faults direction. For a long time (say 1000 years); due to large deformation of ground surface and direction of faults the time for migration of gas to ground surface will be decreased (Figure 8). More analysis related to failure, principle stresses and evolution of gas pressure as well as the former cases can be done.

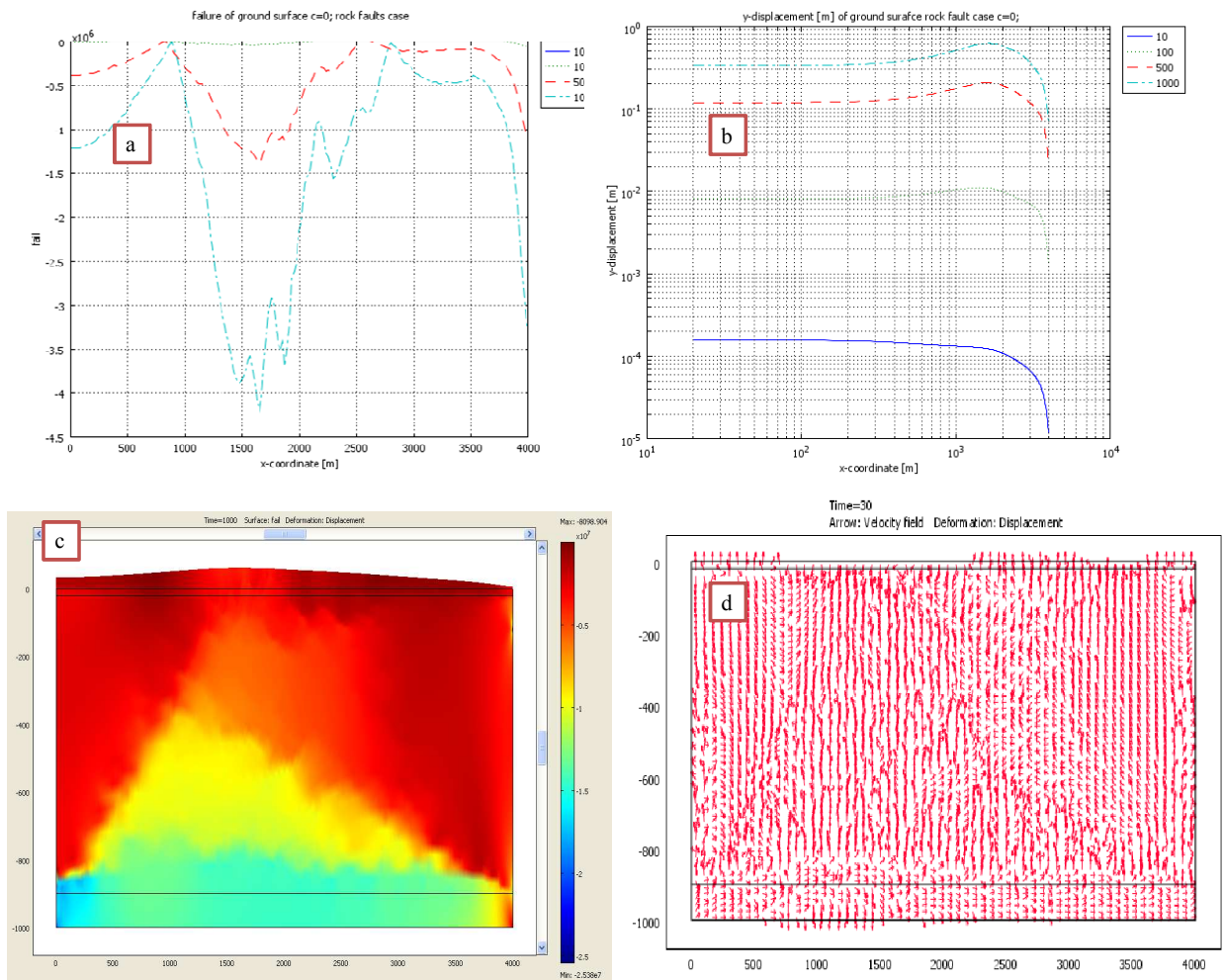

**Figure 7.** a) value of failure for ground surface (second case) during different time scales after injection of CO2; b) vertical displacements for ground surface for 10 to 1000 years ;c) the most probable areas to fail and collapsing after 1000 years and d) direction of gas flow within the faults (faulty zone)

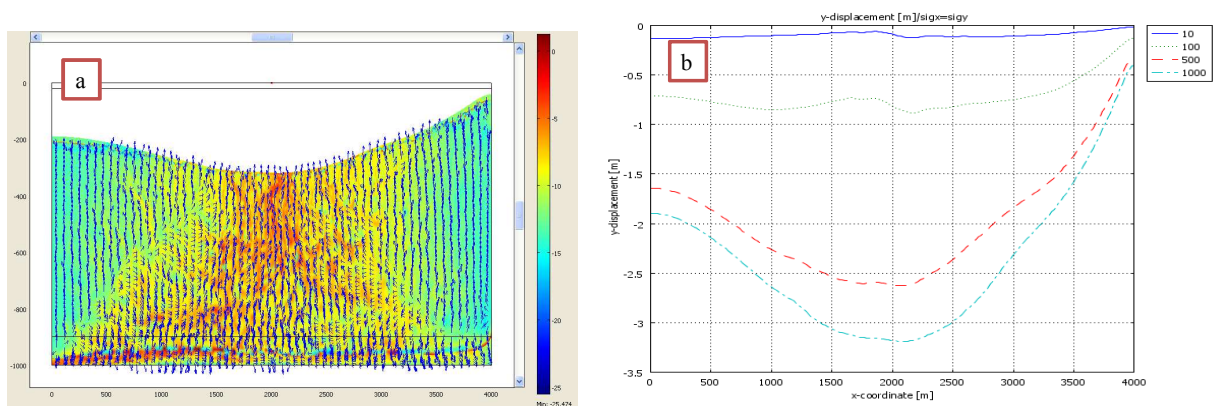

**Figure 8.** a) Logarithmic scale of gas velocity after 1000 years plus Deformation of ground surface due to consolidation when simultaneously the gas is injected (model with considering active stress field) and b) ground surface deformation (subsidence) .

## 4. Conclusion

The effects of heterogeneity to migration of injected CO2 using a hydro-mechanical model and based on a poro-mechanics theory was modeled. Also, over different cases and based on week coupling the evolution of the formation due to gas injection with low rate was investigated. In the first basic case the homogenous case was considered. Second case covered the effects of directional heterogeneity, constructed by random faults, on the flow paths of gas and other attributes of the system. Also, in the latter case the impact of stress state as an active loads (body loads) was regarded. Thus, distinguishable gas flow paths were recognized. In another process, the failure ability of system regard to Mohr-Columb criterion was measured. The results over different cases showed absedince of ground surface (heave), more probable propagation of failure area and the role of directional heterogeneity to change the evolution path of system.